\documentclass[10pt,journal]{IEEEtran}

\usepackage{cite}

\ifCLASSINFOpdf
\else
  \usepackage[dvips]{graphicx}
  \DeclareGraphicsExtensions{.eps}
\fi
\usepackage[cmex10]{amsmath}
\usepackage{amsfonts,amssymb,amsthm,mdwmath}
\usepackage{array}

\ifCLASSOPTIONcompsoc
  \usepackage[caption=false,font=normalsize,labelfont=sf,textfont=sf]{subfig}
\else
  \usepackage[caption=false,font=footnotesize]{subfig}
\fi

\usepackage{stfloats}
\usepackage{url}
\usepackage{color}
\usepackage{acronym}

\begin{document}
\acrodef{PPP}[PPP]{Poisson Point Process}
\acrodef{CDF}[CDF]{Cummulative Distribution Function}
\acrodef{PCF}[PCF]{Pair Correlation Function}
\acrodef{PDF}[PDF]{Probability Distribution Function}
\acrodef{RV}[RV]{Random Variable}
\acrodef{i.i.d.}[i.i.d.]{independent and identically distributed}

\title{Temporal Correlation of Interference in Vehicular Networks with Shifted-Exponential Time Headways}

\author{Konstantinos Koufos and Carl P. Dettmann 
\thanks{K.~Koufos and C.P.~Dettmann are with the School of Mathematics, University of Bristol, BS8 1TW, Bristol, UK. \{K.Koufos, Carl.Dettmann\}@bristol.ac.uk} \protect \\ 
\thanks{This work was supported by the EPSRC grant number 
EP/N002458/1 for the project Spatially Embedded Networks. All underlying data are provided in full within this paper.}}

\maketitle

\begin{abstract}
We consider a one-dimensional vehicular network where the time headway (time difference between successive vehicles as they pass a point on the roadway) follows the shifted-exponential distribution. We show that neglecting the impact of shift in the deployment model, which degenerates the distribution of vehicles to a Poisson Point Process, overestimates the temporal correlation of interference at the origin. The estimation error becomes large at high traffic conditions and small time-lags.
\end{abstract}


\section{Introduction}
The temporal correlation of interference in wireless networks is related to the temporal correlation of outage~\cite{Gong2014,Koufos2018a}, the diversity gain~\cite{Haenggi2013a}, the amount of time a node remains isolated~\cite{Haenggi2013b}, etc., thus, it is an important quantity to study. Interference becomes correlated when it originates from the same set of transmitters~\cite{Haenggi2009}, and the link gains of interfering channels, for some of the transmitters, are correlated over sequential periods of time~\cite{Schilcher2011}. Mobility randomizes the link gains  and naturally decreases interference correlation~\cite{Gong2014,Koufos2018a}. 

The performance of vehicular networks has so far been studied using simplified spatial models, which cannot be considered realistic under all circumstances. Due to its analytical tractability, the \ac{PPP} has been used to model the locations of vehicles along a roadway~\cite{Gong2014,Blaszczyszyn2013}. However, it is known from transportation research that the headway distance (distance between the head of a vehicle and the head of its successor~\cite{Manual2000}, or inter-vehicle distance) depends on the traffic status and it is not always exponential~\cite{Cowan1975,Yin2009}. 

The motivation for this letter is to study the temporal correlation of interference with mobility, considering a more realistic deployment model than the \ac{PPP}. The simplest enhancement shifts the exponential distribution for the inter-vehicle distances to the right. The shift takes into account, to some extent, the interactions between successive vehicles by avoiding unrealistically small headways. We will show that the \ac{PPP} overestimates the temporal correlation of interference in a high traffic scenario. This may affect in return other performance metrics, e.g.,  the conditional probability of success (conditioning on successful reception at the current time slot, the probability to receive successfully also in the next) which would be different than the one predicted by \ac{PPP}. This may further impact the design of retransmission schemes. 

\section{System model}
\begin{figure}[!t]
 \centering
  \includegraphics[width=3.5in]{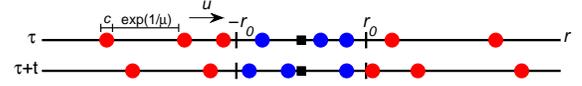}
 \caption{System model illustration at two time slots with time-lag $t$. The vehicles are modeled as identical impenetrable disks. They move rightwards with constant speed $u$. Only the vehicles outside of the cell (red disks) generate interference at the base station (black square). In the figure, the tracking distance is illustrated equal to the diameter of the disk.}
 \label{fig:SystModel2}
\end{figure}
Let us assume that the headway distance consists of a constant tracking distance $c\!>\! 0$ (with probability one) and a free component following the exponential distribution with mean $\mu^{-1}$. The tracking distance models the fact that two vehicles do not come arbitrarily close to each other. In a single lane road with no  overtaking, they are separated at least by the length of a vehicle plus a safety distance. Let us also assume that all vehicles travel with the same constant speed $u$ in the same direction. This might be the case in the flow of convoy or when the vehicles move with the speed limit. This deployment model degenerates to the time headway model M2 proposed by Cowan~\cite{Cowan1975}, i.e., the time headways follow a shifted-exponential distribution. Time-independent statistics of the interference for this model were considered in~\cite{Koufos2018b}.

We study interference correlation at time instances $\tau$ and $\left(\tau\!+\!t\right)$, separated by the time-lag $t$. The base station is located at the origin, and the vehicles located in $[-r_0,r_0]$ are associated to it. The rest generate interference, see Fig.~\ref{fig:SystModel2}. The interferers may communicate with each other in ad hoc mode or paired with other base stations. We would like to get a preliminary insight into the impact of correlated user locations on the temporal aspects of interference at the origin; incorporating further modeling details, e.g., power control and cell association~\cite{Andrews2013}, has been left as a future topic to  study. 

\begin{figure}[!t]
 \centering
  \includegraphics[width=3.2in]{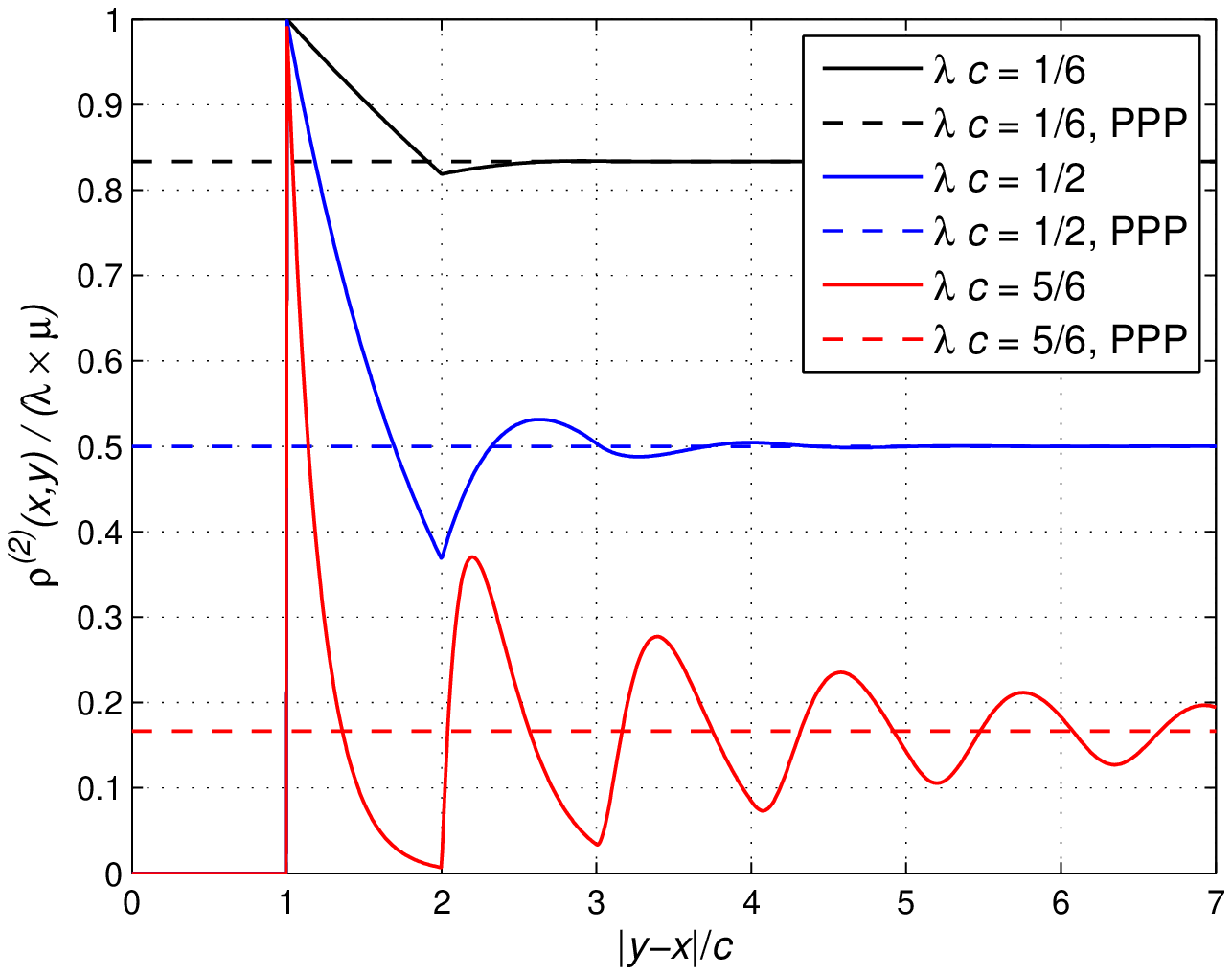}
 \caption{Normalized \ac{PCF}  $\rho^{\left(2\right)}\!\left(x,y\right)\left(\lambda \mu\right)^{-1}$ with respect to the normalized distance $|y-x|c^{-1}$.  The dashed lines correspond to $\rho^{\left(2\right)}\!\left(x,y\right)\!=\!\lambda^2$, or, $\rho^{\left(2\right)}\!\left(x,y\right)\left(\lambda \mu\right)^{-1}\!=\!1\!-\!\lambda c$~\cite[Fig.2]{Koufos2018b}.}
 \label{fig:CorrFunc}
\end{figure}

The \ac{PCF} for a point process where the inter-point distances follow the shifted-exponential distribution has long been studied in the context of statistical mechanics for hardcore fluids/gases~\cite{Salsburg1953} under the name radial distribution function. The point process is stationary. For two vehicles $x,y$ with $y\!>\!x\!:\! y\!\in\!\left(x\!+\!kc,x\!+\!\left(k\!+\!1\right)c\right), k\!\in\!\mathbb{N}$ the \ac{PCF} is depicted in Fig.~\ref{fig:CorrFunc} and has the following form~\cite{Salsburg1953}
\begin{equation}
\label{eq:rho2}
\rho_k^{\left(2\right)}\!\left(y,x\right) = \lambda \sum\limits_{j=1}^k \frac{\mu^j \left(y\!-\!x\!-\!jc\right)^{j-1}}{\Gamma\!\left(j\right) e^{\mu\left(y-x-jc\right)}}, \, k\!\geq\! 1, 
\end{equation}
where $\Gamma\!\left(j\right)\!=\!\left(j\!-\!1\right)!$ is the Gamma function for an integer argument, and $\lambda\!=\!\frac{\mu}{1+\mu c}$ is the intensity of vehicles~\cite{Cowan1975}. 

The distance-based propagation pathloss is $g\!\left(r\right)\!=\! \left|r\right|^{-\eta}$ for $\left|r\right|\!>\!r_0$ and zero otherwise (to filter out vehicles inside the cell), where $\eta\!>\! 2$ is the pathloss exponent. The fast fading $h$ over each link is Rayleigh. Its impact on the interference level is modeled by an exponential \ac{RV} with mean unity. The fading samples from different vehicles are independent \acp{RV}. The transmit power level is unity. 

\section{Correlation coefficient of interference}
For a stationary point process, the mean, $\mathbb{E}\!\left\{\mathcal{I}\right\}$, and the variance, $\mathbb{V}{\text{ar}}\left\{\mathcal{I}\right\}$, of interference are independent of time. The Pearson correlation coefficient at time-lag $t$ is 
\begin{equation}
\label{eq:rho}
\rho\!\left(t\right) = \frac{{\text{cov}}\!\left(\mathcal{I}\!\left(t\right)\right)}{\mathbb{V}{\text{ar}}\left\{\mathcal{I}\right\}} = \displaystyle \frac{ J\!\left(t\right)+ I\!\left(t\right) - \mathbb{E}\!\left\{\mathcal{I}\right\}^2} {\mathbb{V}{\text{ar}}\left\{\mathcal{I}\right\}},
\end{equation}
where ${\text{cov}}\!\left(\mathcal{I}\!\left(t\right)\right)$ is the covariance of interference, and $J\!\left(t\right), I\!\left(t\right)$ are the contributions due to the movement of a single vehicle and a pair of distinct vehicles respectively.  

The terms $J\!\left(t\right), I\!\left(t\right)$ can be written as an  expectation of summation over the point process $\Phi$ of vehicles. After taking into account the independent and unit-mean fading distributions we get 
\[
\begin{array}{ccl}
J\!\left(t\right) \!\!\! &=& \!\!\! \displaystyle \mathbb{E}\!\left\{ \sum\limits_{x\in\Phi} \!\! g\!\left(x\right) g\!\left(x\!+\!tu\right)\right\} \\
I\!\left(t\right) \!\!\! &=& \!\!\! \displaystyle \mathbb{E}\!\left\{ \sum\limits_{x,y\in\Phi}^{x\neq y} \!\! g\!\left(x\right) g\!\left(y\!+\!tu\right) \right\}.
\end{array}
\]

Due to the Campbell's Theorem for stationary processes, the mean interference, $\mathbb{E}\!\left\{\mathcal{I}\right\}$, is equal to the mean interference due to a \ac{PPP} of same intensity, $\mathbb{E}\!\left\{\mathcal{I}\right\}\!=\!2\lambda\!\int_{r_0}^\infty\!g\!\left(r\right)\!{\rm d}r\!=\!\frac{2\lambda r_0^{1-\eta}}{\eta-1}$. In~\cite{Koufos2018b}, we have shown how to approximate the variance of interference, $\mathbb{V}\text{ar}\!\left\{\mathcal{I}\right\}$, due to the point process $\Phi$ and relate it to the variance due to a \ac{PPP} of  intensity $\lambda$. The study in~\cite{Koufos2018b} is about a single time slot. On the other hand, in this letter, we show how to approximate the terms $J\!\left(t\right), I\!\left(t\right)$ and study the behaviour of interference over time.
\begin{figure}[!t]
 \centering
  \includegraphics[width=3.2in]{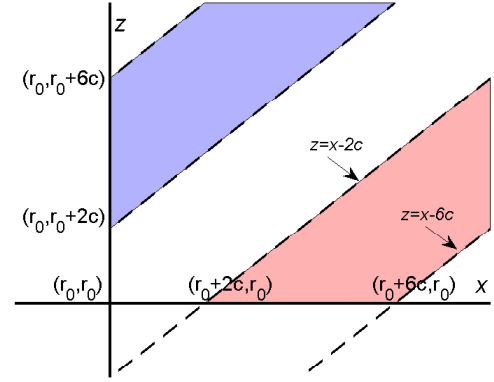}
 \caption{$I_3\!+\!I_4\!=\!\int_{r_0}^\infty\!\!\int_{x-6c}^{x-2c}  g\!\left(x\right)g\!\left(z\right){\rm d}z {\rm d}x \!\stackrel{(a)}{=}\! \int_{r_0}^\infty\!\!\int_{z+2c}^{z+6c}  g\!\left(x\right)g\!\left(z\right){\rm d}x {\rm d}z \!\stackrel{(b)}{=}\! \int_{r_0}^\infty\!\!\int_{x+2c}^{x+6c}  g\!\left(x\right)g\!\left(z\right){\rm d}z {\rm d}x\!=\!I_1\!+\!I_2$, where $(a)$ follows from changing the integration order, see red-shaded area, and $(b)$ from symmetry.}
 \label{fig:IntDomain}
\end{figure}

We calculate the correlation coefficient for time-lags $t\!\leq\!\frac{2r_0}{u}$, where the correlation is exptected to be high. For this range of values, we do not get contributions to the covariance from vehicles performing handover twice. We start with the term $J\!\left(t\right) \!=\! \lambda \int\! g\!\left(r \right) g\!\left(r\!+\!tu \right) {\rm d}r$, where the contributions come only from the vehicles not performing handover within $t$.
\begin{equation}
\label{eq:SmallTau}
\begin{array}{ccl}
J\!\left(t\right) \!\!\!&=&\!\!\! \displaystyle \lambda\!\!\int_{-\infty}^{-r_0}\!\!\!\!\!\! g\!\left(r\right) \! g\!\left(r\!-\!tu\right) \! {\rm d}r \!+\! \lambda\!\!\int_{r_0}^{\infty}\!\!\!\!\!\! g\!\left(r\right) g\!\left(r+tu\right) \! {\rm d}r  \\ \!\!\!&\stackrel{(a)}{=}&\!\!\! \displaystyle \frac{2\lambda r_0^{1-2\eta}}{2\eta-1} {}_2F_1\!\left(2\eta-1,\eta,2\eta,-\frac{t u}{r_0}\right), 
\end{array}
\end{equation}
where $(a)$ follows from setting $x\!=\! - r$ in the first integral, and ${}_2F_1$ is the Gaussian hypergeometric function~\cite[pp.~556]{Abramo}.  

We continue with the contributions to the covariance due to pairs. Let us assume that at time $\tau$, one vehicle is located at the infinitesimal interval ${\rm d}x$, centered at $x$, and the other at the infinitesimal interval ${\rm d}y$, centered at $y$. Since the speed $u$ is constant, the displacement within $t$ is deterministic. After writing $I\!\left(t\right)$ as an integral and using the \ac{PCF} in~\eqref{eq:rho2}, we get
\begin{equation}
\label{eq:It}
\begin{array}{ccl}
I\!\left(t\right) \!\!\!\!\!&=&\!\!\!\!\! \displaystyle \int \!\!  g\!\left(x\right)g\!\left(y+tu\right) \rho^{\left(2\right)}\!\left(x,y\right) {\rm d}y {\rm d}x \\ 
\!\!\!\!\!&=&\!\!\!\!\! \displaystyle  \lambda\!\sum\limits_{k=1}^\infty \!\frac{\mu^k}{\Gamma\!\left(k\right)}\!\! \int\limits_{r_0}^\infty\!\!\!\!\Bigg(\!\!\!\int\limits_{x+kc}^\infty \!\!\!\!\!\! g\!\left(x\right)\!g\!\left(y\!+\!tu\right) \! \frac{\left( y\!-\!x\!-\!kc\right)^{\!k-\!1}\!\!{\rm d}y\!}{e^{\mu\left( y-x-kc\right)}} +\!\! \\ 
{} \!\!\!\!\!& &\!\!\!\!\!\!\!\!\!\!\!\! \displaystyle \int_{-\infty}^{x-kc} \!\!\!\! g\!\left(x\right)\!g\!\left(y\!+\!tu\right)  \frac{\left( x\!-\!y\!-\!kc\right)^{k-1}}{e^{\mu\left( x-y-kc\right)}} {\rm d}y \Bigg){\rm d}x \, + \\ 
{} \!\!\!\!\!& &\!\!\!\!\!\!\!\!\!\!\!\!\!\! \displaystyle \lambda\!\sum\limits_{k=1}^\infty \frac{\mu^k}{\Gamma\!\left(k\right)} \!\! \int\limits_{-\infty}^{-r_0}\!\!\Bigg(\!\!\! \int\limits_{x+kc}^\infty \!\!\!\!\!\! g\!\left(x\right)\!g\!\left(y\!+\!tu\right) \! \frac{\left( y\!-\!x\!-\!kc\right)^{\!k-\!1}\!\!{\rm d}y\!}{e^{\mu\left( y-x-kc\right)}} +\!\! \\ 
{} \!\!\!& & \!\!\!\!\!\!\!\!\!\! \displaystyle \int_{-\infty}^{x-kc} \!\!\!\!\!\!\!\! g\!\left(x\right)\!g\!\left(y\!+\!tu\right)  \frac{\left( x\!-\!y\!-\!kc\right)^{k-1}}{e^{\mu\left( x-y-kc\right)}} {\rm d}y \Bigg){\rm d}x. 
\end{array}
\end{equation}

Equation~\eqref{eq:It} does not provide much insight about the impact of traffic parameters, $\lambda, c$, on the covariance. To get that, we will assume a small tracking distance $c$ as compared to the mean inter-vehicle distance $\lambda^{-1}$. 

\section{Approximation for the covariance}
\label{sec:covariance}
For $\lambda c\!\ll\! 1$, we may approximate the \ac{PCF} with the \ac{PCF} of \ac{PPP} for distance separation larger than $2c$, without introducing much error, $\rho^{\left(2\right)}\!\left(y,x\right) \!\approx\! \lambda^2, \left|y-x\right|\!>\!2c$. For small $\lambda c$, it is the random part  dominating the deployment. Having a vehicle at $x$ imposes little constraint on the probability of finding a vehicle at $y$, given that $x$ and $y$ are far apart. This justifies why for small $\lambda c$, the \ac{PCF}  converges at few multiples of $c$ to the \ac{PCF} of \ac{PPP}, see Fig.~\ref{fig:CorrFunc}. According to~\eqref{eq:It}, the pairs separated by more than $2c$ give a contribution to the term $I\!\left(t\right)$ which can be written as a sum of four terms $I_{>2c}\!\left(t\right)\!=\!\sum\nolimits_{j=1}^4 I_j$.
\[
\begin{array}{ccl}
I_1 \!\!\!&\triangleq&\!\!\! \displaystyle  \lambda^2\!\!\int_{r_0}^\infty\!\!\int_{x+2c}^\infty\!\!\!\!\!g\!\left(x\right)g\!\left(y\!+\!tu\right){\rm d}y {\rm d}x.
\end{array}
\] 

For the term $I_2$, we have to separate between the vehicles at the left- and the right-hand side of the cell at  $\left(\tau\!+\!t\right)$. 
\[ 
\begin{array}{ccl}
I_2  \!\!\!\!\!&\triangleq&\!\!\!\!\! \displaystyle \lambda^2\!\!\int_{r_0}^\infty\!\int_{-\infty}^{x-2c}\!\!\!\!\!g\!\left(x\right)g\!\left(y\!+\!tu\right){\rm d}y {\rm d}x \\ \!\!\!\!\!&=&\!\!\!\!\! \displaystyle  \!\lambda^2\!\!\int_{r_0}^\infty\!\!\!\bigg( \!\int_{-\infty}^{-r_0-tu}\!\!\!\!\!\!\!\!\!\!\!\!\!\!\!\!\!\!g\!\left(x\right)g\!\left(y\!+\!tu\right)\!{\rm d}y \!+\!\int_{r_0-tu}^{x-2c} \!\!\!\!\!\!\!\!\!\!\! g\!\left(x\right)g\!\left(y\!+\!tu\right)\!{\rm d}y\!\bigg) \!{\rm d}x \\ 
\!\!\!\!\!&\stackrel{(a)}{=}&\!\!\!\!\! \displaystyle \!\lambda^2\!\!\int_{r_0}^\infty\!\!\bigg(\!\int_{-\infty}^{-r_0}\!\!\!\!\!\!\!g\!\left(x\right)g\!\left(z\right)\!{\rm d}z \!+\!\int_{r_0}^{x+tu-2c} \!\!\!\!\!\!\!\!\!\!\!\!\!\!\! g\!\left(x\right)g\!\left(z\right)\!{\rm d}z\bigg) \!{\rm d}x \\ \!\!\!\!\!&=&\!\!\!\!\! \displaystyle \!\lambda^2\!\!\int_{r_0}^\infty\!\!\bigg(2\!\int_{r_0}^{\infty}\!\!\!g\!\left(x\right)g\!\left(z\right)\!{\rm d}z \!-\!\int_{x+tu-2c}^\infty \!\!\!\!\!\!\!\!\!\!\!\!\!\!\! g\!\left(x\right)g\!\left(z\right)\!{\rm d}z\bigg) \!{\rm d}x \\ \!\!\!\!\!&=&\!\!\!\!\! \displaystyle 
\frac{1}{2} \mathbb{E}\!\left\{\mathcal{I}\right\}^2 - \lambda^2\!\! \int_{r_0}^\infty\!\!\int_{x+tu-2c}^\infty \!\!\!\!\!\!\!\!\!\!\! g\!\left(x\right)g\!\left(z\right){\rm d}z {\rm d}x,
\end{array}
\]
where $(a)$ follows from $z\!=\!y+tu$ in both integrals and the assumption that $t\!\geq\!\frac{2c}{u}\!\triangleq\!t_1$ in the second. 

After changing the variable $z\!=\!y\!+\!tu$ also in $I_1$, and using the above form for $I_2$, the sum $\left(I_1\!+\!I_2\right)$ can be read as 
\[ 
\begin{array}{ccl}
I_1\!+\!I_2 \!\!\!\!\!\!& = &\!\!\!\!\!\! \displaystyle \frac{1}{2} \mathbb{E}\!\left\{\mathcal{I}\right\}^2 - \lambda^2 \int_{r_0}^\infty\!\!\int_{x+tu-2c}^{x+tu+2c} \!\!\!\!\!\!\!\!\!\!\! g\!\left(x\right)g\!\left(z\right){\rm d}z {\rm d}x, \, t\!\geq\!t_1.
\end{array}
\]

\noindent 
The sum $\left(I_3\!+\!I_4\right)$ can be simplified in a similar manner 
\[ 
\begin{array}{ccl}
I_3\!+\!I_4 \!\!\!\!\!\!& = &\!\!\!\!\!\! \displaystyle \frac{1}{2} \mathbb{E}\!\left\{\mathcal{I}\right\}^2 - \lambda^2 \int_{r_0}^\infty\!\!\int_{x-tu-2c}^{x-tu+2c} \!\!\!\!\!\!\!\!\!\!\! g\!\left(x\right)g\!\left(z\right){\rm d}z {\rm d}x, \, t\!\geq\!t_1.
\end{array}
\]

For time-lags, $t\!\leq\!\frac{2r_0-2c}{u}\!\triangleq\!t_2$, the lower limit in the above integral is larger than $-r_0$, and we get $I_1\!+\!I_2\!=\!I_3\!+\!I_4$. This can be seen by swapping the order of integration in the calculation of $\left(I_3\!+\!I_4\right)$, and keeping in mind that $g\!\left(z\right)\!=\!0,\, \left|z\right|\!\leq\!r_0$, see Fig.~\ref{fig:IntDomain} for an illustration with $tu\!=\!4c$. Finally, for $t\!\in\!\left[t_1,t_2\right]$, the term $I_{>2c}\!\left(t\right)$ becomes 
\begin{equation}
\label{eq:IBigger2c}
\begin{array}{ccl}
I_{>2c}\!\left(t\right) \!\!\!\!\!& = &\!\!\!\!\! \displaystyle  \mathbb{E}\!\left\{\mathcal{I}\right\}^2 - 2 \lambda^2 \int_{r_0}^\infty\!\int_{x+tu-2c}^{x+tu+2c} \!\!\!\!\!\!\!\!\!\!\! x^{-\eta} z^{-\eta} {\rm d}z {\rm d}x \\ \!\!\!\!\!& = &\!\!\!\!\!  \mathbb{E}\!\left\{\mathcal{I}\right\}^2 \!\!+\! \frac{r_0^{2-2\eta} \lambda^2}{\left(\eta-1\right)^2}\!\bigg(\! {}_2F_1\!\left(\!2\eta\!-\!2,\eta,2\eta\!-\!1,-\frac{2c+tu}{r_0}\!\right) - \\ \!\!\!\!\!& &\!\!\!\!\!\!\!\!\!\!\!\!\!\!\!\!\!\!\!\!\!\!\!\!\!\!\!\!\!\   {}_2F_1\!\left(\!2\eta\!-\!2,\eta,2\eta\!-\!1,\frac{2c-tu}{r_0}\!\right)\!\bigg) + \frac{2r_0^{2-2\eta} \lambda^2}{\left(2\eta\!-\!1\right)\left(\eta-1\right)} \bigg(\!\frac{2c-tu}{r_0} \times \\ \!\!\!\!\!&  &\!\!\!\!\!\!\!\!\!\!\!\!\!\!\!\!\!\!\!\!\!\!\!\!\!\!\!\!\!\!\!\!\!\!\!\! {}_2F_1\!\left(\!2\eta\!-\!1,\!\eta,\!2\eta,\!\frac{2c-tu}{r_0}\!\right)\!+\!\frac{2c+tu}{r_0} {}_2F_1\!\left(\!2\eta\!-\!1,\!\eta,\!2\eta\!-\!1,\!-\frac{2c+tu}{r_0}\!\right)\!\!\bigg)\!.
\end{array}
\end{equation}

\noindent 
After expanding~\eqref{eq:IBigger2c} for $b\!=\!\frac{c}{r_0}\!\rightarrow\! 0$ we get 
\begin{equation}
\label{eq:ItBigger2cSmall}
I_{>2c}\!\left(t\right) \approx \mathbb{E}\!\left\{\mathcal{I}\right\}^2 - \frac{8\lambda^2 c \, r_0^{1\!-\!2\eta}}{2\eta-1}\, {}_2F_1\!\left(2\eta\!-\!1,\eta,2\eta,\!-\frac{tu}{r_0}\right)\!.
\end{equation}

It remains to calculate the contribution to the covariance due to pairs of vehicles at distances less than $2c$. For this range of distances, we  consider the exact \ac{PCF}. The term $I_{<2c}\!\left(t\right)$ can  also be written as a sum of four integrals, $I_{<2c}\!\left(t\right)\!=\!\sum\nolimits_{j=4}^8 I_j$. 
\begin{equation}
\label{eq:I5}
\begin{array}{ccl}
I_5 \!\!\!\!\!&\triangleq&\!\!\!\!\!  \displaystyle \lambda\mu\!\!\int_{r_0}^\infty\!\!\!\int_{x+c}^{x+2c}\!\! \frac{g\!\left(x\right)g\!\left(y\!+\!tu\right)}{e^{\mu\left(y-x-c\right)}}{\rm d}y {\rm d}x \\ \!\!\!\!\!&\stackrel{(a)}{=}&\!\!\!\!\! \displaystyle \lambda \mu^{\eta}\!\!\! \int_{r_0}^\infty\!\!\!\!\! x^{-\eta} e^w \big( \Gamma\!\left(1\!-\!\eta,\!w\right)\!-\! \Gamma\!\left(1\!-\!\eta,\!w\!+\!\mu c\right) \! \big){\rm d}x, 
\end{array}
\end{equation}
where in $(a)$ we changed the variable $z\!=\!y\!+\!tu$ before integrating and then, we substituted   $w\!=\!\mu\left(x\!+\!tu\!+\!c\right)$.  $\Gamma\!\left(a,x\right)\!=\!\int_x^\infty t^{a-1} e^{-t} {\rm d}t$ is the incomplete Gamma function.

Due to the fact that $x\!\geq\!r_0\!$ and $\mu\!=\!\frac{\lambda}{1-\lambda c}\!\geq\!\lambda$, when there are, on average,  many vehicles inside the cell, or equivalently $\lambda r_0\!\gg\!1$, we may approximate the integrand in~\eqref{eq:I5} for large $w$. After approximating up to $w^{-1}$ order we have 
\[
\begin{array}{ccl}
I_5 \!\!\!&\approx&\!\!\! \displaystyle \lambda\left(1\!-\!e^{-c\mu}\right)\!\! \int_{r_0}^\infty\!\!\! x^{-\eta} \left(x\!+\!tu\!+\!c\right)^{-\eta}{\rm d}x \, - \\ \!\!\!& &\!\!\! \displaystyle \frac{\eta\lambda\left(1\!-\!e^{-c\mu}\!-\!c\mu e^{-c\mu}\right)}{\mu}\!\!\int_{r_0}^\infty\!\!\! x^{-\eta} \left(x\!+\!tu\!+\!c\right)^{-\eta-1} {\rm d}x \\  
\!\!\!&=&\!\!\!  \lambda r_0^{-2\eta}\!\bigg( \frac{\left(1-e^{-c\mu}\right)r_0}{2\eta-1} {}_2F_1\!\left(\eta,2\eta\!-\!1,2\eta,-\frac{c+tu}{r_0}\right) \, + \\ \!\!\!& &\!\!\! \frac{c\mu e^{-c\mu}-1+e^{-c\mu}}{2\mu} {}_2F_1\!\left(2\eta,\eta\!+\!1,2\eta\!+\!1,-\frac{c+tu}{r_0}\right)\!\bigg).
\end{array}
\]

\noindent 
The term $I_6$ is 
\[
\begin{array}{ccl}
I_6 \!\!\!&\triangleq&\!\!\!  \displaystyle \lambda\mu\!\int_{r_0}^\infty\!\!\!\int_{x-2c}^{x-c}\!\!\frac{g\!\left(x\right)g\!\left(y\!+\!tu\right)}{e^{\mu\left(x-y-c\right)}}{\rm d}y {\rm d}x.
\end{array}
\]

For $t\!\geq\!t_1$, the vehicle located at ${\rm d}y$ at time $\tau$ will always generate interference at time $\left(\tau\!+\!t\right)$. Therefore we can follow similar steps used to approximate  $I_5$.
\[
\begin{array}{ccl}
I_6  \!\!\!&\stackrel{(a)}{=}&\!\!\! \displaystyle \lambda\mu \!\int_{r_0}^\infty\!\!\!\int_{x-2c+tu}^{x-c+tu}\!\!\!\!\!\!\!\!\!x^{-\eta}z^{-\eta} e^{-\mu\left(x-z-c+tu\right)}{\rm d}z {\rm d}x \\ 
\!\!\!&\stackrel{(b)}{=}&\!\!\! \displaystyle \lambda \left(-\mu\right)^\eta \!\! \int_{r_0}^\infty \!\!\!\!\!\!\! x^{-\eta} e^{-w}\left(\Gamma\!\left(1\!-\!\eta,\!-w\right) \!-\! \Gamma\!\left(1\!-\!\eta,\!\mu c\!-\!w\right) \right) {\rm d}x \\ \!\!\!&\stackrel{(c)}{\approx}&\!\!\! \displaystyle \lambda \mu^\eta \!\!\int_{r_0}^\infty \!\!\!\!\!\!x^{-\eta} w^{-\eta} \left(1\!-\!e^{-c\mu}\!+\!\frac{e^{c\mu}\left(1\!+\!c\mu\right)\!-\!2c\mu\!-\!1}{e^{c\mu} w}\right)\! {\rm d}x \\ 
\!\!\!&=&\!\!\! \lambda r_0^{-2\eta}\!\bigg( \frac{\left(1-e^{-c\mu}\right)r_0}{2\eta-1} {}_2F_1\!\left(\eta,2\eta\!-\!1,2\eta,\frac{c-tu}{r_0}\right) \, + \\ \!\!\!& &\!\!\! \frac{1-c\mu e^{-c\mu}-e^{-c\mu}}{2\mu} {}_2F_1\!\left(2\eta,\eta\!+\!1,2\eta\!+\!1,\frac{c-tu}{r_0}\right)\!\!\bigg), 
\end{array}
\]
where $(a)$ follows from $z\!=\!y\!+\!tu$, $(b)$ from integrating in terms of $z$ and then substituting  $w\!=\!\mu\left(x\!+\!tu\!-\!c\right)$, and $(c)$ from expanding the integrand up to $w^{-1}$ order. 

Following similar steps used to show that $I_1\!+\!I_2\!=\!I_3\!+\!I_4$, it is also possible to show that $I_5\!=\!I_7$ and $I_6\!=\!I_8$ for $t\!\in\!\left[t_1,t_2\right]$. After using the above approximations for $I_5$ and $I_6$ into $I_{<2c}\!\left(t\right) = 2\left(I_5+I_6\right)$, substituting $\mu\!=\!\frac{\lambda}{1-\lambda c}$, and expanding up to second-order for small $\lambda c$, we end up with  
\[ 
\begin{array}{ccl}
I_{<2c}\!\left(t\right)  \!\!\!\!\!&\approx&\!\!\!\!\! \frac{\lambda^2 c \,  r_0^{1-2\eta}}{2\left(2\eta-1\right)} \bigg(\! 2\left(2\!+\!c\lambda\right)\Big( {}_2F_1\!\left(\eta,2\eta\!-\!1,2\eta,b\!-\!\frac{tu}{r_0}\right) + \\ \!\!\!\!\!\!\!\!\!\!\!\!& &\!\!\!\!\!\!\!\!\!\!\!\!\! {}_2F_1\!\left(\eta,2\eta\!-\!1,2\eta,-b\!-\!\frac{tu}{r_0}\right)\!\!\Big) \!+\! b\left(2\eta\!-\!1\right) \times \\ \!\!\!\!\!\!\!\!\!\!\!\!\!\!\!\!\!\!\!\!\!\!\!\!\!\!\!& &\!\!\!\!\!\!\!\!\!\!\!\!\!\!\!\!\!\!\!\!\!\!\!\!\!\!\!\!\! \left({}_2F_1\!\left(\!2\eta,\!\eta\!+\!1,\!2\eta\!+\!1,\!b\!-\!\frac{tu}{r_0}\!\right) \!-\! {}_2F_1\!\left(\!2\eta,\!\eta\!+\!1,\!2\eta\!+\!1,\!-b\!-\!\frac{tu}{r_0}\!\right)\!\right) \!\!\! \bigg).
\end{array}
\]
\begin{figure}[!t]
 \centering
  \includegraphics[width=3.2in]{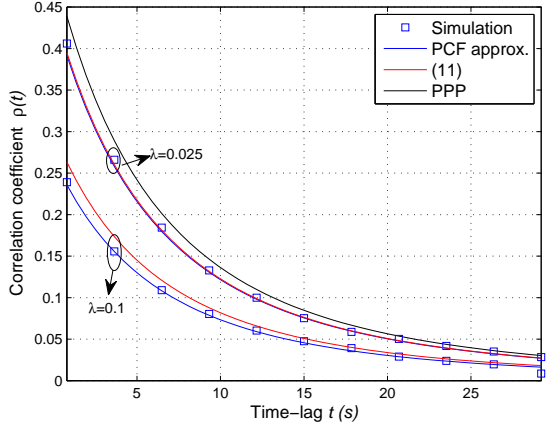}
 \caption{Correlation coefficient of interference. $10^5$ simulations. Pathloss exponent $\eta\!=\!3$, cell size $r_0\!=\! 150$ m, tracking distance $c\!=\! 4$ m, speed $u\!=\! 10$ m/s. For the '\ac{PCF} approx.',   $I_{>2c}\!\left(t\right)$ is calculated from~\eqref{eq:IBigger2c}, and $I_{<2c}\!\left(t\right)\!=\! 2\lambda\mu\!\!\int_{r_0}^\infty\!\!\left( \!\int_{x+c}^{x+2c} \frac{g\!\left(x\right)g\!\left(y+tu\right)}{e^{\mu\left(y-x-c\right)}}{\rm d}y \!+ \!\int_{x-2c}^{x-c}\frac{g\!\left(x\right)G\left(y+tu\right)}{e^{\mu\left(x-y-c\right)}}{\rm d}y \right)\! {\rm d}x$ is integrated numerically.}
 \label{fig:RhoLam}
\end{figure}

\noindent 
After expanding $I_{<2c}\!\left(t\right)$ for $b\!=\! \frac{c}{r_0}\!\rightarrow\! 0$ we have 
\begin{equation}
\label{eq:ItSmaller2cSmall}
\begin{array}{ccl}
I_{<2c}\!\left(t\right)  \!\!\!\!\!&\approx&\!\!\!\!\! \displaystyle \frac{2\lambda^2 c \left(2\!+\!\lambda c\right)r_0^{1-2\eta} {}_2F_1\!\left(\!2\eta-1,\!\eta,\!2\eta,\!-\frac{tu}{r_0}\!\right)}{2\eta-1}. 
\end{array}
\end{equation}

Finally, we substitute~\eqref{eq:SmallTau},~\eqref{eq:ItBigger2cSmall} and~\eqref{eq:ItSmaller2cSmall} into the covariance, cancel out the means and do some factorization. 
\begin{equation}
\label{eq:covsmall}
\begin{array}{ccl}
\text{cov}\!\left(\mathcal{I}\!\left(t\right)\right) \!\!\!\!\!&\approx&\!\!\!\!\! \displaystyle \frac{2 \lambda r_0^{1-2\eta}{}_2F_1\!\left(\!2\eta\!-\!1,\!\eta,\!2\eta,\!-\!\frac{tu}{r_0}\!\right)}{2\eta-1} \!\left(1 \!-\! \lambda c\right)^2.  
\end{array}
\end{equation}

\noindent 
For $c\!=\!0$,~\eqref{eq:covsmall} degenerates to the covariance of \ac{PPP}. 

\section{Approximation for the correlation coefficient}
The temporal correlation coefficient of interference for a \ac{PPP} is  independent of the intensity of vehicles $\lambda$. After substituting  $c\!=\!0$ in~\eqref{eq:covsmall}, and using that the variance of interference for a \ac{PPP} is $\frac{4\lambda r_0^{1-2\eta}}{2\eta-1}$, we calculate 
\[
\rho_{\text{PPP}}\!\left(t\right)  = \frac{1}{2}\, {}_2F_1\!\left(2\eta-1,\eta,2\eta,-\frac{t u}{r_0}\right)\!.
\]

In order to approximate the correlation coefficient for the point process $\Phi$, we also need an approximation for the variance of interference. This has been derived in~\cite[Eq.(13)]{Koufos2018b} 
\begin{equation}
\label{eq:VarianceHC}
\mathbb{V}{\text{ar}}\left\{\mathcal{I}\right\} \approx \frac{4\lambda r_0^{1-2\eta}}{2\eta-1} \left(1-\lambda c \right). 
\end{equation}

After substituting~\eqref{eq:covsmall}$-$\eqref{eq:VarianceHC} in~\eqref{eq:rho}, we can capture the impact of intensity and tracking distance on the correlation coefficient. 
\begin{equation}
\label{eq:RhoSmall}
\begin{array}{ccl}
\rho\!\left(t\right) \!\!\!&\approx&\!\!\! \displaystyle  \left(1-\lambda c\right) \rho_{\text{PPP}}\!\left(t\right). 
\end{array}
\end{equation}

The approximation in~\eqref{eq:RhoSmall} indicates that the \ac{PPP} overestimates the temporal correlation of interference. The error is more prominent for increasing $\lambda c$ and small time-lags. This is  illustrated in Fig.~\ref{fig:RhoLam}. We see over there that for both values of $\lambda$, using the approximation for the \ac{PCF} at distances larger than $2c$ introduces negligible error. The error of~\eqref{eq:RhoSmall} is mostly due to the expansions around $\lambda c\!\rightarrow\! 0$ in the approximation of the variance in~\eqref{eq:VarianceHC} and of the term $I_{<2c}\!\left(t\right)$ in the covariance.

A more realistic model for the location of vehicles along a roadway indicates a lower correlation of interference over time. In order to capture the impact of lower correlation on the receiver performance, a relevant metric could be the conditional probability of success or outage. For that, we need models for the outage probability over a single time slot and also over two slots, i.e., a bivariate distribution model with correlated marginals. Since the probability generating functional of the point process with shifted-exponential inter-arrivals is unknown, this is an interesting topic for future work.

\section{Conclusions}
The \ac{PPP} allows unrealistically small headways. Avoiding this by shifting the exponential distribution for the inter-vehicle distances to the right, reduces the temporal correlation of interference. For fixed tracking distance $c$, the interference might not be that correlated, particularly in high traffic conditions, large $\lambda$. The performance assessment of a receiver at the origin, e.g., temporal outage probability, local delay, etc., and the mechanisms to cope with the high correlation predicted by the \ac{PPP} model may need to be revisited. Future topics of study may include more realistic headway models and may also incorporate more accurate models for the uplink.

\end{document}